\documentclass{anabs}
\usepackage{graphicx}
\usepackage{times}
\Volume{00}              
\Year{0000}              
\Month{00}               
\Pagespan{000}{000}      

\begin{document}
\lhead[\thepage]{A.N. Morales-Rueda: The LT Integral Field Unit}
\rhead[Astron. Nachr./AN~{\bf XXX} (200X) X]{\thepage}
\headnote{Astron. Nachr./AN {\bf 32X} (200X) X, XXX--XXX}

\title{The Liverpool Telescope Spectrograph: FRODOSpec}

\author{L. Morales-Rueda$^1$, D. Carter$^2$, I. A. Steele$^2$,
  P. A. Charles$^1$, S. Worswick$^3$}
\institute{1. University of Southampton, UK, 2. Liverpool John Moores
  University, UK, 3. Observatory Optics, 1 Betoni Vale, UK}
\date{Received {date will be inserted by the editor}; 
accepted {date will be inserted by the editor}} 

\abstract{We describe in some detail one of the instruments that will
be available in 2004 to the research community. FRODOSpec is an
integral field unit spectrograph that will be available for use with
the 2\,m robotic Liverpool Telescope on the island of La Palma. We
anticipate that this instrument will open up major areas of research
that cannot be carried out with conventionally operated
telescopes. Some of these research areas relate to indirect imaging of
astronomical objects like Doppler tomography and eclipse mapping.
\keywords{} }

\correspondence{lmr@astro.soton.ac.uk}

\maketitle

\section{The Liverpool Telescope \&\ FRODOSpec}

The Liverpool Telescope (LT) is a 2\,m fully robotic Cassegrain
telescope sited at the Observatorio del Roque de los Muchachos in La
Palma. Three main instruments are being built for use at this
telescope: an optical CCD camera, a near infra-red camera and an
integral field unit, the Fibre-fed RObotic Dual-beam Optical
Spectrograph (FRODOSpec).

FRODOSpec's front end consists of a lenslet array with each lenslet
fed to an optical fibre. The output fibres are lined up in a
pseudo-slit and passed through a dichroic beam-splitter, allowing us
to separate the blue and red regions of the spectrum. The optics that
follow are then optimised for each wavelength region thereby
maximising the total throughput. Each beam or arm contains two
different dispersing elements giving 2 resolutions in the red and two
in the blue. The final result is an intermediate resolution
spectrograph (R$\sim$4000) that covers the full optical wavelength
range from 3800 to 9500\AA, with a higher resolution option
(R$\sim$8000) covering two wavelength ranges, 3800 -- 5600 and 5600 --
7400\AA.

FRODOSpec, with an integral field unit field of view of 11 x 11
arcsecs, has been designed mainly to study point sources. The robotic
nature of the telescope plus spectrograph will allow us to monitor
objects over a wide range of temporal baselines. FRODOSpec will be
able to achieve a time resolution between 5 and 10 s. We will also
make use of the target of opportunity advantage that a robotic
telescope can give, e.g. this will allow us to study dwarf novae in
their outburst state which is the step needed for understanding the
viscous evolution of their accretion discs and thereby constrain the
nature of that viscosity. The third advantage of a robotic
spectrograph is to permit queue scheduling, whereby observations for a
particular project will take place when the conditions are optimal,
e.g. given orbital phase, seeing.

Time-lapse spectroscopy by FRODOSpec will allow us to use Doppler
Tomography to perform indirect imaging of interacting binaries on the
microarcsecond scale. We are also considering the addition of
polarimetric capabilities to the spectrograph in the future. This
would allow observers to perform Stokes Imaging and Zeeman Doppler
Tomography. For 1-dimensional indirect imaging techniques like Eclipse
Mapping and Physical Parameter Eclipse Mapping, observers can obtain
colour photometry with the CCD camera.

\begin{figure}
\resizebox{\hsize}{!}
{\includegraphics[angle=-90]{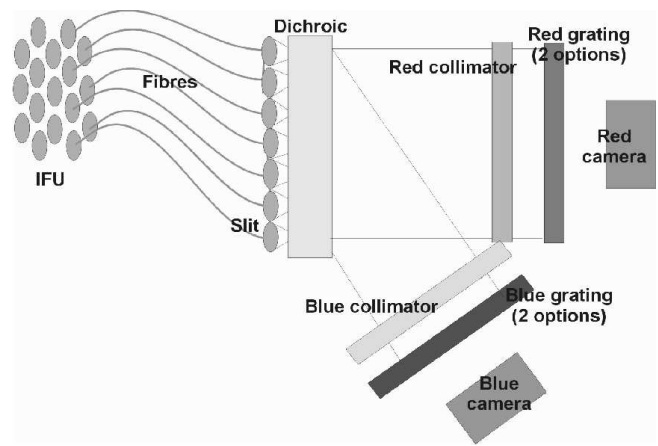}}
\caption{Schematic layout for FRODOSpec.}
\label{figlabel}
\end{figure}

\section{Useful facts}

The breakdown of the available time on the LT is as follows:\\ 70\%
for the United Kingdom (of which 30\% is for John Moores University),
20\% for Spanish astronomers, 5\% for international applicants and the
remaining 5\% for education and public understanding of science in the
UK. Observing time will be allocated by PATT in the UK, CAT in Spain
and CCI for the international time. Successful applicants will be given
fully reduced data. The observing time is assigned in hours and the
switch between imaging and spectroscopy is effected via the
positioning of a deployable tertiary mirror. The data will be public
and stored in an archive one year after the observations for a given
programme are completed.

FRODOSpec is being developed jointly by Southampton and Liverpool John
Moores Universities. For more information on the LT and FRODOSpec check
the following website http://telecope.livjm.ac.uk.



\end{document}